\documentclass[3p,times,10pt]{elsarticle}
\usepackage{amsfonts}
\usepackage{amssymb}
\usepackage{amsmath}
\usepackage{graphicx}
\usepackage{epsfig}
\usepackage{ctable}

\begin{document}

\begin{frontmatter}

\title{Quantum transport in coupled resonators enclosed synthetic magnetic flux}
\author{L.~Jin\corref{cor1}}
\ead{jinliang@nankai.edu.cn}
\cortext[cor1]{Corresponding author}
\address{School of Physics, Nankai University, Tianjin 300071, China}

\begin{abstract}
Quantum transport properties are instrumental to understanding quantum coherent transport processes. Potential applications of quantum transport are widespread, in areas ranging from quantum information
science to quantum engineering, and not restricted to quantum state
transfer, control and manipulation. Here, we study light transport
in a ring array of coupled resonators enclosed synthetic magnetic flux.  The ring configuration, with an arbitrary number of resonators embedded, forms an two-arm Aharonov-Bohm interferometer. The influence of magnetic flux on light transport is investigated. Tuning the magnetic flux can lead to resonant transmission, while half-integer magnetic flux quantum leads to completely destructive interference and transmission zeros in an interferometer with two equal arms.
\end{abstract}


\end{frontmatter}

\section{Introduction}

\label{sec_introduction} Coherent transport in discrete low-dimensional
mesoscopic quantum systems has attracted much attentions for fundamental
physics as well as applications due to persistent technique progress in
nanofabrication~\cite{Schuster,Holleitner,Johnson,Chakrabarti10}. Quantum
interference phenomena and quantum coherent transport properties are
extensively investigated in various quantum devices~\cite{Datta},
interesting effects including Fano resonance~\cite{Clerk,Fano,Chakrabarti07}, Kondo effect~\cite{Kondo1,Kondo2}, Aharonov-Bohm (AB) effect~\cite{AB,Chakrabarti11}, have been theoretically predicted and experimentally verified in one-dimensional nanostructures system, such as photonic
nanocrystals~\cite{NanoCrystal1,NanoCrystal2,NanoCrystal3}, nanowires~\cite{nanowire1,nanowire2}, quantum dots array~\cite{QDA,QDA2}. Much effort has been devoted to investigate coherent electron transport properties of a various types of interferometers oriented at different geometries~\cite{singledot,Iye,Joea,Bichler,Kim}. Among these, quantum dots embedded two-arm electron wave interferometers are mostly studied, in particular, the AB interferometer~\cite{Holleitner,Kim,ABHolleitnerScience}.
The AB interferometer is composed by a two-terminal structure with single or double quantum dots embedded. The two-terminal configuration forms a close ring shape quantum system threading by magnetic flux that is tunable via externally applied magnetic field. Studying transport properties of a quantum system is beneficial for understanding quantum processes concerning energy transfer. The
applications include quantum state transfer, quantum control and
manipulation in quantum information science.

The engineered and modulated optical systems mimic a great deal of quantum
systems including atomic, molecular in condensed matter physics. For example, the
dynamic localization \cite{DL} and Bloch oscillation \cite{BO} of
electron under static external field are demonstrated in periodically
modulated coupled optical waveguides. The optical systems of coupled
waveguides and coupled resonators are successfully employed to investigate parity-time symmetry~\cite{PT,AGuo,Ruter,Peng}. In recent years, quantum
optical analogue is a fruitful platform for studying coherent mechanics
in quantum realm due to the simplicity and flexibility of photons. Photons, as neutral particles, although proved not directly interact with magnetic field, the photonic analogy of AB effect is proposed~\cite{ArtificialGaugeField,ArtificialGaugeField2} and verified based on dynamic modulation of material permittivity~\cite{Fang2012PRL,Fang2012}, and based on photon-phonon interaction~\cite{Li2013}. The key point is to realize nonreciprocal photon tunneling. Thereafter, the study of quantum phenomena from solids to their artificial electromagnetic counterparts in optics has become a hot topic in quantum physics, quantum information and quantum optics~\cite{Fang2012,Li2013,Fang2013,Carusotto,Tzuang2014,Lu2014}. The synthetic artificial gauge field is introduced to simulate and investigate phenomena usually found in electronic systems~\cite{Carusotto}, investigations include the quantum topology in photonic system~\cite{Hafezi2011,Hafezi2013,Hafezi2014,Mittal2014} and the photonic analogue of quantum (anomalous) Hall effect in photonic crystals~\cite{Joannopoulos2008,Wang2008,Wang2009,Ozawa2014}.

Here, we study light transport in the presence of synthetic magnetic flux in a ring array of coupled resonators. In the presence of artificial gauge field, photons mimic electrons in magnetic field, a coupled resonators system provides an optical platform for investigating quantum phenomena. The ring array of coupled resonators enclosed synthetic magnetic field acts as a two-arm AB interferometer. As interference is sensitive to the relative phase between light waves propagating in its arms, thus light transmission probability is relevant to the connection positions of the input and output leads, and depends on the enclosed synthetic magnetic flux. The interference and phase sensitive feature reveal the quantum nature of photons. We employ the Bethe ansatz method to study the light transport properties. We focus on the influence of a synthetic gauge field on light transport in a ring shaped array with arbitrary number of coupled resonators. The transmission zeros and the resonant transmissions are further discussed.

\section{Coupled resonators enclosed synthetic magnetic flux}

\begin{figure}[t]
\centering
\includegraphics[clip,width=16 cm]{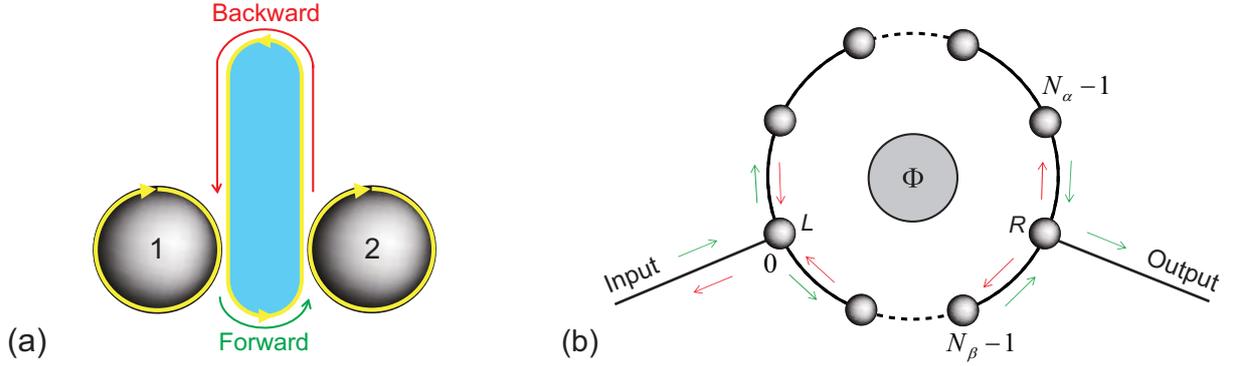}
\caption{(a) The coupled resonators. Two primary ring resonators $1$, $2$ coupled
together via an antiresonant auxiliary resonator (in blue). The optical paths (green
and red arrows) imbalance induces a nonreciprocal hopping phase between two primary resonators. (b) The ring array of coupled resonators enclosed synthetic magnetic flux. The synthetic magnetic flux is introduced through nonreciprocal hoppings between neighbor resonators based on (a). The input lead is connected to resonator $L$, the output lead is connected to resonator $R$. The system acts as a two-arm AB ring interferometer. The output light can be modulated by the enclosed synthetic magnetic flux $\Phi $.}
\label{fig1}
\end{figure}
We begin with two coupled passive resonators~\cite{Ramezani}, using coupled mode theory, the dynamics of field amplitudes inside the resonators give \begin{eqnarray}
\frac{\mathrm{d}a_{1}}{\mathrm{d}t} &=&-i\omega _{c}a_{1}+iJa_{2}, \\
\frac{\mathrm{d}a_{2}}{\mathrm{d}t} &=&-i\omega _{c}a_{2}+iJa_{1},
\end{eqnarray}%
where $a_{1}$, $a_{2}$ are the field amplitude in the two resonators, $%
\omega _{c}$ is the resonant frequency, $J$ is the coupling strength. The
coupled resonators system is able to be described by a tight-binding
Hamiltonian, $H_{0}=-J\hat{a}_{1}^{\dagger }\hat{a}_{2}-J\hat{a}%
_{2}^{\dagger }\hat{a}_{1}$. Similarly, an $N$ coupled passive resonators
system with resonant frequency $\omega _{c}$ and coupling strength $J$ is
described by $H_{\mathrm{ring}}=-J\sum_{j=1}^{N}(\hat{a}_{j}^{\dagger }\hat{%
a}_{j+1}+\hat{a}_{j+1}^{\dagger }\hat{%
a}_{j})$, where $\hat{a}_{j}^{\dagger }$ ($\hat{a}_{j}$ ) is
the photon creation (annihilation) operator for the $j$-th resonator with
periodical boundary condition that $\hat{a}_{j+N}^{\dagger }=\hat{a}%
_{j}^{\dagger }$, the roundtrip length of each resonator is $L=n\lambda $,
where $\lambda $ is the light wavelength, $n$ is an integer.

The synthetic magnetic flux is induced by nonreciprocal hopping phase
introduced in coupled resonators as proposed by Hafezi~\cite%
{Hafezi2014}. For any two adjacent resonators, the hopping is induced
through an auxiliary resonator (Fig.~\ref{fig1}a). When optical length of the auxiliary resonator is $L^{\prime }=n\lambda +(3/2)\lambda $, being antiresonant with two primary resonators, the auxiliary resonator results in an effective coupling strength $J$ in between two primary resonators describing by the Hamiltonian $H_0$. In Fig.~\ref{fig1}a, light in the resonators is shown in yellow. The forward going path of light from resonator $1$ to $2$ is shown in green, the backward going path of light from resonator $2$ to $1$ is shown in red. Considering the path length difference between the forward and backward direction is $\Delta x_0$, the effective coupling acquires an additional optical path dependent hopping phase factor $e^{i\phi _{0}}$ with $\phi
_{0}=2\pi \Delta x_{0}/\lambda $. The auxiliary resonator between two primary resonators induces a nonreciprocal coupling $H_{\rm{eff}}=-Je^{i\phi
_{0}}\hat{a}_{1}^{\dagger }\hat{a}_{2}-Je^{-i\phi _{0}}\hat{a}%
_{2}^{\dagger }\hat{a}_{1}$. For a ring configuration shown in Fig.~\ref{fig1}b based on the building blocks shown in Fig.~\ref{fig1}a, photons circling around the ring array accumulate an AB type phase factor $e^{-i\Phi }$ ($e^{+i\Phi }$) in clockwise (counterclockwise) direction, the total phase is $\Phi
=\sum_{j}\phi _{j}$. The effective AB type phase introduced in the ring array equivalently realizes a synthetic magnetic flux for photons. The synthetic magnetic field felt by photons is gauge invariant and unaffected by the distribution of nonreciprocal phase as long as the total phase $\Phi $ accumulated by photons in one round is unchanged. The ring resonator supports clockwise and counterclockwise modes, here we assume clockwise mode (yellow circles) as shown in Fig.~\ref{fig1}a. The situation for counterclockwise mode is similar, where photons feel an exactly opposite synthetic magnetic flux.

To be precise, we consider a ring array with $N$ coupled resonators threading by synthetic magnetic flux $\Phi$. The system is schematically illustrated in Fig.~\ref{fig1}b, each site (black ball) represents a ring resonator. The input and output leads connect to the array of coupled resonators at two different resonators ($L$ and $R$). The connection leads split the ring array into two components: The upper arm and the lower arm. The ring array of coupled resonators functions as a two-arm AB interferometer. The light can propagate from input resonator ($L$) to output resonator ($R$) either in the upper or the lower arm. Photons tunneling in the upper arm acquire a difference phase comparing with tunneling in the lower arm, meet at the output resonator and interferes with each other. The light interference depends on the different phases photons acquired in the two propagating paths (the upper and the lower ring arms). Therefore, the light output is affected by the synthetic magnetic flux, which alternates the relative phase difference between two optical paths. In this sense, the wave vector of light and the connection positions of the input and output leads also influence the light transmission. In the following, we investigate light transport properties in a ring array of coupled resonators in the framework of interference. The Hamiltonian of a two-arm ring interferometer under synthetic magnetic flux is written as
\begin{equation}
H=-J\sum_{j=1}^{N_{\alpha }}e^{i\phi }\hat{a}_{\alpha ,j-1}^{\dagger }\hat{a}%
_{\alpha ,j}-J\sum_{j=1}^{N_{\beta }}e^{-i\phi }\hat{a}_{\beta ,j-1}^{\dagger
}\hat{a}_{\beta ,j}+\mathrm{H.c.,}
\end{equation}%
where $\hat{a}_{\alpha ,j}^{\dagger }$ ($\hat{a}_{\alpha ,j}$ ) is the
creation (annihilation) operator of resonator $j$ in the upper ring arm.
Corresponding creation and annihilation operators for resonator $j$ of the lower
ring arm are denoted by $\hat{a}_{\beta ,j}^{\dagger }$ and $\hat{a}_{\beta
,j}$. In order to investigate the influence of synthetic gauge field in the system, the coupling strength $J$ is set unity, the resonators are the same in resonant frequency $\omega_c$. There are no disorders in the couplings and no detunings in the array of coupled resonators, the light transport is solely controlled and affected by the synthetic magnetic flux and the ring interferometer structure. We consider the input lead is connected to the ring array at resonators $0$ and $N_{\alpha }$ ($N_{\beta }$, which is actually resonator $N_{\alpha }$). The threading synthetic magnetic flux in the ring array is $\Phi =(N_{\alpha }+N_{\beta })\phi $, it acts globally and is invariant under local operator transformations.

\section{Interference in coupled resonators and light transport}

We study the dynamical process by employing the Bethe ansatz method. Equivalent descriptions of a ring array of coupled resonators is guaranteed under
local operator transformation. For convenience, we take a local transformation $\hat{a}%
_{\alpha ,j}^{\dagger }\rightarrow e^{-i\phi j}\hat{a}_{\alpha ,j}^{\dagger
} $ with $j\in \lbrack 1,N_{\alpha }]$ in the upper ring arm, and $\hat{a}_{\beta ,j}^{\dagger
}\rightarrow e^{i\phi j}\hat{a}_{\beta ,j}^{\dagger }$ with $j\in \lbrack
1,N_{\beta }-1]$ in the lower ring arm. The Hamiltonian is expressed as
\begin{equation}
H=-\sum_{j=0}^{N_{\alpha}-1}\hat{a}_{\alpha ,j}^{\dagger }\hat{a}_{\alpha
,j+1}-\sum_{j=0}^{N_{\beta }-2}\hat{a}_{\beta ,j}^{\dagger }\hat{a}_{\beta
,j+1}-e^{-i\Phi }\hat{a}_{\beta ,N_{\beta }-1}^{\dagger }\hat{a}_{\beta
,N_{\beta }}+\mathrm{H.c.,}
\end{equation}
which describes a one-dimensional coupled ring resonators system with effective nonreciprocal tunneling phase introduced only in between resonators $N_{\beta-1}$ and $N_{\beta}$. Photons acquire an additional phase $\Phi$ tunneling from resonator $N_{\beta-1}$ to $N_{\beta}$, and acquire an additional phase $-\Phi$ when tunneling inversely from resonator $N_{\beta}$ to $N_{\beta-1}$.

The supporting light frequency in the ring array of coupled resonators is in the range of $\omega \in [\omega_{c}-2J, \omega_{c}+2J]$. As $J$ is set unity, we consider an incident light wave with frequency $\omega=\omega_{c}-2\mathrm{cos}k$ ($\pi\leq k\leq\pi$) incoming in the left lead and outgoing in the right lead. The ring array of coupled resonators splits the input wave at connection resonator $L$, a portion of light reflects back to the input lead and others propagate forward in the ring arms; The forward propagating lights meet at resonator $R$, interfere with each other and form the output. Meanwhile, the interference at output resonator $R$ also results in a portion of light reflecting back toward input resonator $L$, meeting at the input resonator $L$, interfere and form reflection light in the input lead. In the light propagating process, the ring array of coupled resonators acts as a two-arm AB interferometer. The interference depends on the phase accumulated by photons at each path, which depends on enclosed synthetic magnetic flux in the ring array of coupled resonators.

\begin{figure}[t!]
\centering
\includegraphics[width=16cm]{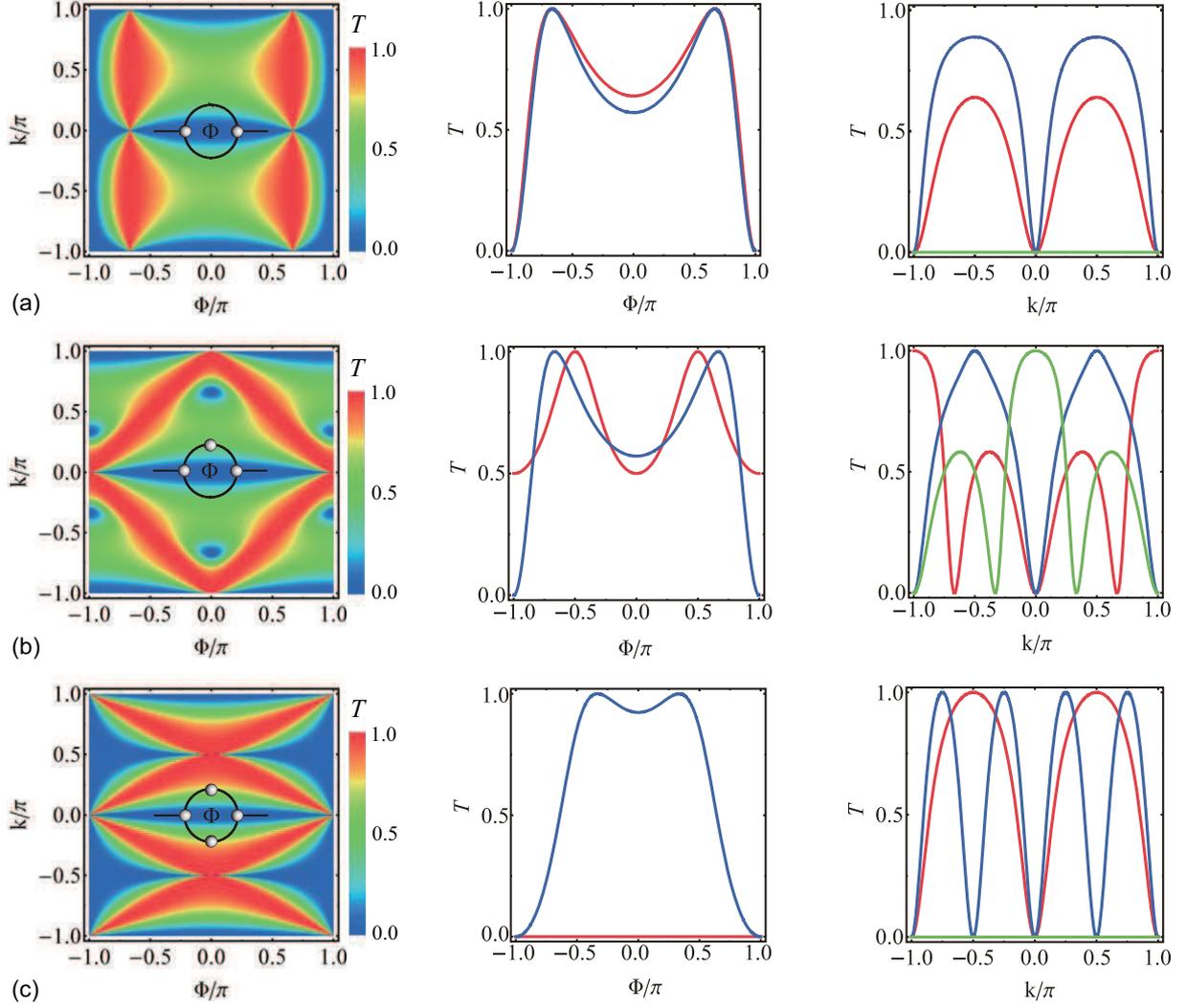}
\caption{The light transmission probability $T$. We schematically illustrate the ring array of coupled resonator in the inserts of (a) for no resonator embedded in two arms, (b) for one resonator embedded in one arm, and (c) for two resonators respectively embedded in two arms. Corresponding contours of $T$ are plotted in the left panels as functions of the enclosed synthetic magnetic flux and the light wave vector. The middle panels are cuts through contours of $T$ at $k=\protect\pi/2$ (red) and $k=\protect\pi/3$ (blue). The right panels are cuts through contours of $T$ at $\Phi=0$ (red), $\protect\pi/2$ (blue), $\protect\pi$ (green).}
\label{fig2}
\end{figure}

The magnetic flux affects periodically, the phase factor $e^{i\Phi}$ is unchanged for every $2\pi$ period of $\Phi$, i.e., $e^{i\Phi}=e^{i\Phi+2n\pi}$ with integer $n\in\mathbb{Z}$. Thus, the light transmission probability in the system satisfies $T(\Phi)=T(\Phi+2n\pi)$. In Fig.~\ref{fig2}, we plot the light transmission probabilities for several different configurations with $\Phi \in [-\pi,\pi]$. Moreover, for the ring array shown in Fig.~\ref{fig1}b, the transmission probabilities are irrelevant to the direction of the magnetic flux. The magnetic flux is opposite when we turn over the ring array, which indicates an exchanging of the upper and the lower arms of the two-arm AB ring interferometer. This will not change the light interference, and the light transmission remains unchanged. Thus, we have $T(\Phi)=T(-\Phi)$, which implies the counterclockwise mode supporting by the coupled resonators system has the same propagating properties as the clockwise mode we are discussing. The ring array of coupled resonators is a Hermitian system, the light transmission probabilities are the same for light inputting in either side, this implies a symmetric light transport $T(k)=T(-k)$.

The dynamics in the system is governed by the Schr\"{o}dinger equations. The field amplitude in each resonator has the form $a=\psi e^{-i\omega t}$. In a rotating frame at oscillating frequency $\omega_c$, the stationary resonator modal amplitude $\psi$ inside the ring array of coupled resonators is set as
\begin{equation}
\psi _{\alpha,j}=X_{\alpha }e^{ikj}+Y_{\alpha
}e^{-ikj},
\end{equation}
\begin{equation}
\psi _{\beta,j}=X_{\beta }e^{ikj}+Y_{\beta }e^{-ikj},
\end{equation}
where $e^{+ikj}$ and $e^{-ikj}$ represent the forward and backward propagating light waves with vector $k$, respectively. The stationary resonator modal amplitude in the leads resonators of the input and output leads are set as $e^{ikj}+re^{-ikj}$ and $te^{ikj}$. The reflection and transmission coefficients are defined as $r$ and $t$ for wave vector $k$. Due to the interference, light waves in the upper and the low arms of the ring array contain forward transmitted (green arrow in Fig.~\ref{fig1}) and backward reflected (red arrow in Fig.~\ref{fig1}) components. In order to calculate the reflection and transmission coefficients, we list the Schr\"{o}dinger equations for all resonators in the whole coupled resonators system. Notice that the Schr\"{o}dinger equations for resonator $j_{\alpha }\in
\lbrack 1,N_{\alpha }-1]$ in the upper arm and resonator $j_{\beta }\in \lbrack
1,N_{\beta }-2]$ in the lower arm give a series of equations,
\begin{equation}
-(X_{\sigma }e^{ik(j_{\sigma }-1)}+Y_{\sigma }e^{-ik(j_{\sigma
}-1)})-(X_{\sigma }e^{ik(j_{\sigma }+1)}+Y_{\sigma }e^{-ik(j_{\sigma
}+1)})=E_{k}(X_{\sigma }e^{ikj_{\sigma }}+Y_{\sigma }e^{-ikj_{\sigma }}),
\label{eq:3}
\end{equation}%
the footnote is $\sigma =\alpha ,\beta$. These equations all yield the eigen energy $E_{k}=-2\mathrm{cos}k$, which implies the dispersion relation, $\omega=\omega _{c}-2\mathrm{cos}k$. The remaining Schr\"{o}dinger equations in the ring array together with the boundary conditions (continuity of wave functions at the connection resonators) determine the light propagating properties. The boundary conditions mean that the modal amplitude $\psi _{\alpha,0}$, $\psi _{\alpha,N_{\alpha} (N_{\beta})}$ should be in accordance at resonator $0$ and $N_{\alpha }$ ($N_{\beta }$). The boundary conditions at resonator $0$ are related to the reflection coefficient,
\begin{equation}
X_{\alpha }+Y_{\alpha }=X_{\beta }+Y_{\beta }=1+r.
\end{equation}%
Similarly, the boundary conditions at resonator $N_{\alpha }$ are related to the transmission coefficient,
\begin{equation}
X_{\alpha }e^{ikN_{\alpha }}+Y_{\alpha }e^{-ikN_{\alpha }}=t.
\end{equation}%
The Schr\"{o}dinger equation at resonator $N_{\beta }-1$ in the lower ring arm gives
\begin{equation}
-(X_{\beta }e^{ik(N_{\beta }-2)}+Y_{\beta }e^{-ik(N_{\beta }-2)})-e^{i\Phi
}t=E_{k}[X_{\beta }e^{ik(N_{\beta }-1)}+Y_{\beta }e^{-ik(N_{\beta }-1)}],
\end{equation}%
this yields another boundary condition at resonator $N_{\beta }$,
\begin{equation}
X_{\beta }e^{ikN_{\beta }}+Y_{\beta }e^{-ikN_{\beta }}=e^{i\Phi }t,
\end{equation}%
the Schr\"{o}dinger equation for resonator $0$ gives%
\begin{equation}
( X_{\alpha }e^{ik}+Y_{\alpha }e^{-ik})+( X_{\beta
}e^{ik}+Y_{\beta }e^{-ik}) =e^{ik}+re^{-ik},
\end{equation}%
and for resonator $N_{\alpha }$ gives
\begin{equation}
[ X_{\alpha }e^{ik(N_{\alpha }-1)}+Y_{\alpha }e^{-ik(N_{\alpha }-1)}] +e^{-i\Phi }[ X_{\beta }e^{ik(N_{\beta }-1)}+Y_{\beta
}e^{-ik(N_{\beta }-1)}] =te^{-ik}.  \label{eq:8}
\end{equation}%
From equations (\ref{eq:3}-\ref{eq:8}) obtained from the Bethe ansatz method, we can exactly calculate the reflection and transmission coefficients $r$, $t$. In the following, we discuss light transmissions in details, in particular, the influence of synthetic magnetic flux.

The situation of $\mathrm{sin}(N_{\alpha }k)=\mathrm{sin}(N_{\beta
}k)=0$ in the ring array induces a completely destructive interference, that input light is totally reflected with zero transmission. This is an interference effect only relevant to the structure and the input wave vector. The physical interpretation of destructive interference is as follows. The forward and backward going waves interfere in the ring array, forming two standing waves in the two arms. The standing waves are sinusoidal functions that exactly being the eigenstates of uniform tight-binding chains with $N_{\alpha }-1$ and $N_{\beta }-1$ coupled resonators, and the interference in the ring array results in vanishing of wave functions at the connection resonators. In this situation, two ring arms that function as transport channels are isolated, and we have $r=-1$, $t=0$. The input light will be fully reflected without any transmission, while the synthetic magnetic flux will not affect the light propagation. In other aspect, the gauge invariant field introduced in open chain is trial. In the situation of $\mathrm{sin}(N_{\alpha }k)=\mathrm{sin}(N_{\beta
}k)=0$, the ring array of coupled resonators will act as a cage constraining photons of wave vector $k$ initially localized in the ring array. Photons travel in the ring arms without escaping to the leads.

For the ring array and input wave vector satisfying $\mathrm{sin}(N_{\alpha
}k)=0$ and $\mathrm{sin}(N_{\beta }k)\neq 0$, the light wave vector satisfies $%
e^{2iN_{\alpha }k}=1$, the phase acquired by photons in the path of upper ring arm from resonator $0$ to resonator $N_{\alpha}$ is $e^{iN_{\alpha}k}$. This indicates the wave functions at two connection resonators have the same amplitude and the same (opposite) phase, which depends on the length of upper ring arm, $e^{iN_{\alpha }k}=1$ ($e^{iN_{\alpha }k}=-1$). The same amplitude of light in the connection resonators implies a none zero transmission. Otherwise, zero transmission will reduce to the situation of $\mathrm{sin}(N_{\alpha }k)=\mathrm{sin}(N_{\beta
}k)=0$. In the situation $\mathrm{sin}(N_{\alpha
}k)=0$ and $\mathrm{sin}(N_{\beta }k)\neq 0$, none zero light transmission can also be seen from the transmission coefficient $t$, which is calculated by the Bethe ansatz method. From Eq. (\ref{eq:3}-\ref{eq:8}), we have
\begin{equation}
t=\frac{i\mathrm{sin}(N_{\beta }k)\mathrm{sin}k}{\mathrm{cos}\Phi \mathrm{sin%
}k+e^{iN_{\alpha }k}\{\mathrm{sin}[(N_{\beta }-1)k]+i\mathrm{sin}(N_{\beta
}k)\mathrm{sin}k\}},  \label{eq:t_caseII}
\end{equation}%
where we directly notice the transmission $\left\vert t\right\vert ^{2}>0$ because $\mathrm{sin}(N_{\beta }k)\neq 0$, and the enclosed synthetic magnetic flux modulates the light transmission probability. The resonant light transmission $\left\vert t\right\vert ^{2}=1$ happens at $\mathrm{%
cos}\Phi =-e^{iN_{\alpha }k}\mathrm{sin}[(N_{\beta }-1)k]/\mathrm{sin}k$. Through tuning the enclosed synthetic magnetic flux in the system, we can always have a resonant light transmission for an input light with wave vector $k=\pi/2$, which corresponds to the input frequency $\omega_c$, i.e., the resonator on-resonance frequency. The on-resonance input is an optimal choice for obtaining resonant light transmission through controlling synthetic magnetic flux. For input wave vector $k\neq \pi/2$, although a synthetic magnetic flux modulates the light transmission, a resonant light transmission $\left\vert t\right\vert ^{2}=1$ may be impossible by controlling the synthetic magnetic flux, e.g. $|\mathrm{sin}[(N_{\beta }-1)k]|>|\mathrm{sin}k|$. The case for $\mathrm{sin}(N_{\alpha }k)\neq 0$ and $\mathrm{sin}(N_{\beta
}k)=0$ is similar, the transmission coefficient obtained is Eq. (\ref%
{eq:t_caseII}) with subscript $\alpha$ and $\beta$ switched ($\alpha \longleftrightarrow \beta $) and the transmission coefficient $t$ substituted by $te^{i\Phi }$ ($t\rightarrow
te^{i\Phi }$).

Now we turn to discuss a general situation. After simplification Eq. (\ref{eq:3}-\ref{eq:8}) under $\mathrm{sin}%
(N_{\alpha }k)\neq 0$ and $\mathrm{sin}(N_{\beta }k)\neq 0$, we obtain the transmission coefficient for a two-arm ring interferometer with input and output leads connected at arbitrary positions, the light transmission coefficient of the ring array of coupled resonators threading by synthetic magnetic flux $\Phi$ is in form of
\begin{equation}
t=\frac{2i\left( \frac{1}{\mathrm{sin}(N_{\alpha }k)}+\frac{e^{-i\Phi }}{%
\mathrm{sin}(N_{\beta }k)}\right) \mathrm{sin}^{2}k}{\left( \frac{1}{\mathrm{%
sin}(N_{\alpha }k)}+\frac{e^{i\Phi }}{\mathrm{sin}(N_{\beta }k)}\right)
\left( \frac{1}{\mathrm{sin}(N_{\alpha }k)}+\frac{e^{-i\Phi }}{\mathrm{sin}%
(N_{\beta }k)}\right) \mathrm{sin}^{2}k-\left( \frac{\mathrm{sin}[(N_{\alpha
}-1)k]}{\mathrm{sin}(N_{\alpha }k)}+\frac{\mathrm{sin}[(N_{\beta }-1)k]}{%
\mathrm{sin}(N_{\beta }k)}-e^{-ik}\right) ^{2}}.
\end{equation}
where the transmission zeros $\left\vert t\right\vert ^{2}=0$ happen only when $\Phi
=0,\pi $ at wave vector $k$ fulfills $\mathrm{sin}(N_{\alpha }k)=\mp \mathrm{%
sin}(N_{\beta }k)$. In other words, the transmission zeros are at wave vector
$k=(2m+1)\pi /(N_{\alpha }-N_{\beta })$ when there is no magnetic flux $\Phi
=0$ and $k=2m\pi /(N_{\alpha }-N_{\beta })$ when the magnetic flux is at
maximum value $\Phi =\pm \pi $, where $m$ is integer number. The wave functions in the ring arms are sinusoidal functions, waves in the
upper and the lower arms interfere destructively due to the extra phase factor $%
e^{i\Phi }$ in the presents of magnetic flux. The destructive interference
effectively opens the ring array at the connection resonator $N_{\alpha}$ ($N_{\beta}$) coupled
to the output lead, and the ring arms act like dangling chains. By tuning
magnetic flux with other system parameters set, we obtain two reflection zeros at
different magnetic flux in a $2\pi $ period. The transmission probability $T=\left\vert
t\right\vert ^{2}$ for several simple configurations of ring array under
magnetic flux are shown in Fig.~\ref{fig2}. The whole dynamical process is
unitary. The reflection is $R=1-T$. The ring array with no resonator embedded (Fig.~\ref{fig2}a), one resonator embedded (Fig.~\ref{fig2}b), and two resonators embedded (Fig.~\ref{fig2}c) embedded are shown. Moreover, we notice that $\Phi =\pm \pi $ results in destructive interference in the AB ring interferometer with equal
length ( $N_{\alpha }=N_{\beta }$). This is because the phase
difference of photons reaching the output lead from the upper and from the
lower arm is $\Phi $ for system with $N_{\alpha }=N_{\beta }$, the magnetic flux induces additional phase factor $e^{i\Phi }=-1$. Therefore, half-integer magnetic flux quantum leads to a destructive interference in the two-arm AB ring interferometer with equal length.
\begin{figure}[tb]
\centering
\includegraphics[width=10 cm]{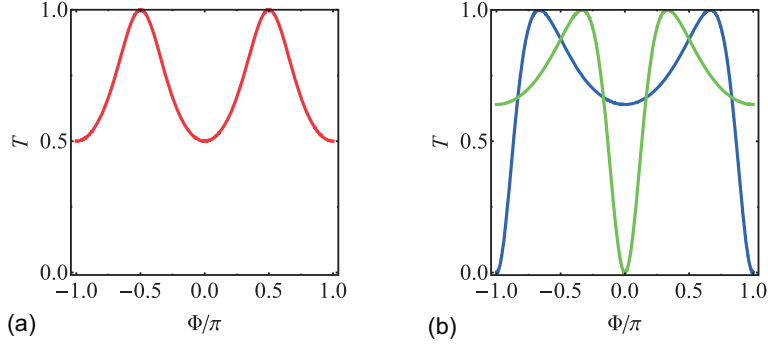}
\caption{The light transmission probability as a function of synthetic magnetic flux for wave vector $k=\protect\pi/2$%
. (a) Either $N_{\protect\alpha}$ or $N_{\protect\beta}$ is odd. (b) Both $%
N_{\protect\alpha}$ and $N_{\protect\beta}$ are odd. The blue line is for system with odd $(N_{\protect\alpha}+N_{\protect\beta})/2$ and green line is for system with even $(N_{\protect\alpha}+N_{\protect\beta})/2$. The transmission is
zero for $N_{\protect\alpha}$ and $N_{\protect\beta}$ being both even.}
\label{fig3}
\end{figure}

The light with wave vector $k=\pi /2$ propagates at the fast velocity $2%
\mathrm{sin}k$, the dispersion relation yields the light frequency is on-resonance with the ring resonators $\omega =\omega _{c}$. Varying the
synthetic magnetic flux enclosed in the ring interferometer, the phase difference photons accumulated in the upper and the lower arms changes, which affects the interference in the ring array. The transmissions probabilities for $k=\pi /2$ is plotted as a function of the magnetic flux $\Phi$ in Fig.~\ref{fig3}. For $k=\pi /2$, the destructive interference induces full reflection without any transmission in case the numbers of coupled resonators embedded in the two interferometer arms ($N_{\alpha }$, $N_{\beta }$) are both even. In the situation that the number of embedded resonators being even in one arm ($N_{\alpha }$ or $N_{\beta }$ is even), the transmission probability equals to $\vert t\vert ^{2}=( \mathrm{cos^{2}\Phi }+1) ^{-1}$, and varies in the range of $0.5$ to $%
1.0$ as the synthetic magnetic flux. The resonant light
transmission happens at $\Phi =\pm \pi /2$, and reaches the minimum $0.5$ at
$\Phi =0$ (Fig.~\ref{fig3}a). For ring array with $N_{\alpha }$, $N_{\beta }$ being both odd, the synthetic magnetic flux induces asymmetric Fano lineshapes in the transmission (Fig.~\ref{fig3}b). In this situation, we have $\vert \mathrm{sin}(N_{\alpha }\pi /2)\vert
=\vert \mathrm{sin}(N_{\beta }\pi /2)\vert =1$, and the transmission
coefficient $t=\left\vert t\right\vert e^{i\varphi _{t}}$ ($\left\vert t\right\vert
$ is the amplitude and $\varphi _{t}$ is the phase) is closely related
to the system structure. If $(N_{\alpha }+N_{\beta })/2$ is odd, we have $(N_{\alpha }-N_{\beta })/2$ being even, therefore $e^{i[(N_{\alpha }-N_{\beta })(\pi/2)-\Phi]}$ equals to $-1$ at half-integer magnetic flux quantum ($\Phi=2n\pi+\pi$, $n\in\mathbb{Z}$), which leads to completely destructive interference (blue line in Fig.~\ref{fig3}b at $\Phi=\pm \pi$). The amplitude is $\left\vert t\right\vert=4\mathrm{cos}(\Phi /2)[4\mathrm{cos}^{2}(\Phi /2)+1]^{-1}$. The phase is $\varphi
_{t}=\pi /2-\Phi /2$ for $\mathrm{sin}(N_{\alpha }\pi /2)=1,\mathrm{sin}%
(N_{\beta }\pi /2)=1$ and $\varphi _{t}=-\pi /2-\Phi /2$ for $\mathrm{sin}%
(N_{\alpha }\pi /2)=-1,\mathrm{sin}(N_{\beta }\pi /2)=-1$, respectively. If $(N_{\alpha
}+N_{\beta })/2$ is even, we have $(N_{\alpha }-N_{\beta })/2$ being odd, therefore $e^{i[(N_{\alpha }-N_{\beta })(\pi/2)-\Phi]}$ equals to $-1$ at integer magnetic flux quantum ($\Phi=2n\pi$, $n\in\mathbb{Z}$), which leads to completely destructive interference (green line in Fig.~\ref{fig3}b at $\Phi=0$). The amplitude is $\left\vert t\right\vert=4\mathrm{sin}(\Phi /2)[4\mathrm{sin}^{2}(\Phi /2)+1]^{-1}$. The phase is $\varphi _{t}=\pi -\Phi /2$ for $\mathrm{%
sin}(N_{\alpha }\pi /2)=1,\mathrm{sin}(N_{\beta }\pi /2)=-1$ and $\varphi
_{t}=-\Phi /2$ for $\mathrm{sin}(N_{\alpha }\pi /2)=-1,\mathrm{sin}(N_{\beta
}\pi /2)=1$, respectively.
\begin{figure}[tb]
\centering
\includegraphics[width=16 cm]{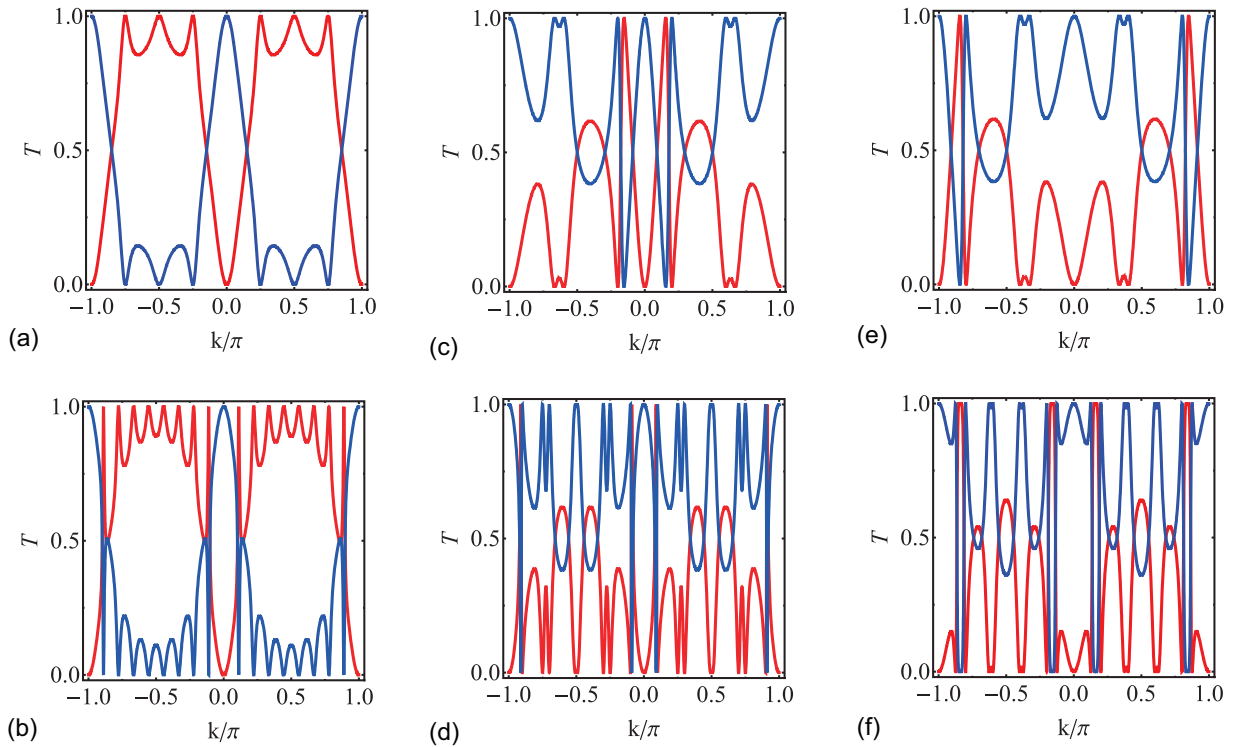}
\caption{The light transmission (red) and reflection (blue) probabilities for system with $N_{%
\protect\alpha }=1$. Other parameters are (a) $N_{\protect\beta }=4,\Phi =\protect\pi /2$; (b) $%
N_{\protect\beta }=9,\Phi =\protect\pi /2$; (c) $N_{\protect\beta }=4,\Phi =%
\protect\pi $; (d) $N_{\protect\beta }=9,\Phi =\protect\pi $; (e) $N_{\protect%
\beta }=4,\Phi =2\protect\pi $; (f) $N_{\protect\beta }=9,\Phi =2\protect\pi $%
.}
\label{fig4}
\end{figure}

In the situation that there is no resonator embedded in the upper arm of the ring array ($N_{\alpha }=1$). The light transmission coefficient for $N_{\beta }$ coupled resonator embedded in the lower arm is
\begin{equation}
t=\frac{2i[ \mathrm{sin(}N_{\beta }k)/\mathrm{sin}k+e^{-i\Phi }]
\mathrm{sin}k}{[ \mathrm{sin(}N_{\beta }k)/\mathrm{sin}k+2\mathrm{cos}%
\Phi ] +2ie^{-iN_{\beta }k}\mathrm{sin}k}.
\label{t_Na1}
\end{equation}%
In Eq. (\ref{t_Na1}), we notice that in the region $k\in (-\pi,0)$ and $k\in (0, \pi)$, there is no transmission zero (%
$\left\vert t\right\vert ^{2}\neq 0$) except for $\Phi =n\pi $, $n\in\mathbb{Z}$ (see Fig. \ref{fig4}). This indicates that only half-integer or integer magnetic flux quantum could lead to complete destructive interference without light transmission. The resonant transmissions ($\left\vert t\right\vert^{2}=1$) happen at $2\mathrm{%
\mathrm{cos}}\Phi =\mathrm{-sin}(N_{\beta }k)/\mathrm{sin}k$. For $\Phi =(n+1)\pi
/2$ ($n\in\mathbb{Z}$), the transmission amplitude is generally larger than $\Phi$ being others and the resonant transmission ($\left\vert t\right\vert ^{2}=1$) happens at $\mathrm{sin}(N_{\beta
}k)=0$. The transmission and reflection are shown in Fig.~\ref{fig4}a and \ref%
{fig4}b for the ring array with $N_{\beta }=4$ and $N_{\beta }=9$, where the resonant transmissions are seen at each $k=m\pi
/N_{\beta }$, $m\in \mathbb{Z}$ for $\Phi =\pi /2$. We also show $\Phi =\pi $
in Fig.~\ref{fig4}c and \ref{fig4}d for $N_{\beta }=4$ and $N_{\beta }=9$,
where the transmission zeros ($\left\vert t\right\vert ^{2}\neq 0$) are at $\mathrm{sin}(N_{\beta }k)=\mathrm{sin}k$. The $\Phi =2\pi $
is a trivial case, the transmission probability is equivalent with no magnetic flux case, we show $\Phi
=2\pi $ in Fig.~\ref{fig4}e and \ref{fig4}f for $N_{\beta }=4$ and $N_{\beta
}=9$ as comparison, where the transmission zeros ($\left\vert t\right\vert
^{2}=0$) are at $\mathrm{sin}(N_{\beta }k)=-\mathrm{sin}k$. The synthetic
magnetic flux modulates the light transport, significantly varies $k$ for the resonant transmission as well as the transmission zeros. This might be useful in the design of quantum devices such as photon filters or optical switches.

\begin{figure}[tbp]
\centering
\includegraphics[width=16cm]{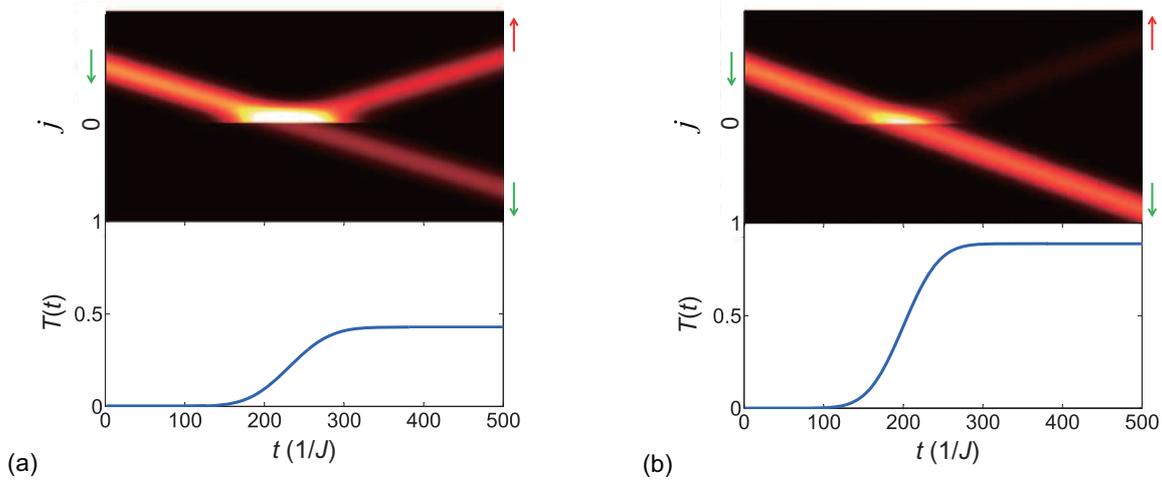}
\caption{The quantum simulation of dynamical transport. The time evolutions of Gaussian
wave packets in a ring array of coupled resonators shown in Fig.~\ref{fig1}b are simulated in the upper panels. The transmission probabilities as a function of time are shown in the lower panels. The system parameters are (a) $N_{\protect\alpha }=3$, $N_{\protect\beta %
}=1,\Phi =0$, $k=\protect\pi /3$; (b) $N_{\protect\alpha }=3$, $N_{\protect%
\beta }=1,\Phi =\protect\pi /2$, $k=\protect\pi /2$.}
\label{fig5}
\end{figure}

To verify our analysis on transmission probability of the ring array of coupled resonators enclosed synthetic magnetic flux, we simulate light transmission through calculating the dynamical evolution of a Gaussian profile wave packet. The profile of the light wave packet is
\begin{equation}
\left\vert G(k,N_{\mathrm{c}})\right\rangle =\Omega
^{-1/2}\sum_{j}e^{-w^{2}(j-N_{c})^{2}/2+ikj} a^{\dagger}_{j}\vert \mathrm{vac} \rangle,
\end{equation}
which is centered at the resonator $N_{\mathrm{c}}$, the coefficient $\Omega^{-1}e^{-w^{2}(j-N_{\mathrm{c}})^{2}}$ represents the renormalized light intensity of resonator $j$, $e^{ikj}$ is the relative phase in each resonators of the incident light wave, and the parameter $w$ controls the width of the Gaussian profile in the array of coupled resonators. The Gaussian wave is localized at its profile center $N_{\mathrm{c}}$ and keeps shape-preserving in the evolution process, in particular, a wide spreading Gaussian wave packet~\cite{Kim2006}. The interference of a plane wave with wave vector $k$ can be approximately
simulated by a Gaussian wave packet with velocity $2J\mathrm{sin}k$ in a uniform chain with nearest neighbor coupling strength $J$. To simulate the dynamics, we consider a ring array of coupled resonators threading by synthetic magnetic flux is the central of the system, connected to the input and output leads. The Gaussian wave with certain velocity is initially localized in the input lead far away from the central, it moves along the input lead towards the central, interferes in the ring array of coupled resonators and outgoes from the output lead. The dynamical process and transmission probability are shown in the upper and the lower panels of Fig.~\ref{fig5} for the Gaussian profile waves in a ring array of coupled resonators system with $N_{\alpha }=3$, $N_{\beta }=1$. Figure~\ref{fig5}a is for wave with $k=\pi /3$ when synthetic magnetic flux is $\Phi =0$, and Fig.~\ref{fig5}b is for wave with $k=\pi /2$ when synthetic magnetic flux is $\Phi =\pi /2$. The numerically simulated transmission probabilities are in accordance with the analytically calculated transmission probabilities, being $3/7$ (Fig.~\ref{fig5}a) and $8/9$ (Fig.~\ref{fig5}b).

\section{Conclusion and discussion}

In conclusion, the influence of synthetic magnetic flux on light transport is investigated for an array of coupled resonators in ring configuration threading by synthetic magnetic flux. The synthetic magnetic flux is introduced in the couplings through optical paths imbalance. We discuss the transport properties of a coupled resonators system with arbitrary number of resonators embedded and arbitrary positions that the leads connected. In particular, the transmission zeros and resonant transmissions are discussed in details. The half-integer magnetic flux quantum results in completely destructive interference and transmission zeros for the ring array of coupled resonators with two arms at equal length, and light can be confined inside without leakage. The ring array of coupled resonators enclosed synthetic magnetic flux can be used to control light transport. Tuning the enclosed magnetic flux, resonant transmission is available, and the sensitivity to synthetic magnetic flux might be useful in the design of photon filters or optical switches. We perform numerical simulations, the results are in accordance with our theoretical analysis. The array of coupled resonators is a discrete quantum system, it also reflects the quantum transport of electrons in magnetic field. Our results might be useful for studying coherent dynamics of a ring Aharonov-Bohm interferometer and may have potential applications in quantum state transfer, quantum control and manipulation in quantum information science.

\section*{Acknowledgment}
This work was supported by National Basic Research Program (973 Program) of China (Grant No. 2012CB921900) and Nankai University Baiqing Plan foundation (Grant No. ZB15006104).

\end{document}